\documentclass[aps,prb,reprint,superscriptaddress,amsmath,longbibliography,amssymb]{revtex4-2}%twocolumn,aps,showpacs,showkeys,prd,superscriptaddress,byrevtex
\usepackage{graphicx}% Include figure files
\usepackage{dcolumn}% Align table columns on decimal point
\usepackage{bm}% bold math
\usepackage{hyperref}
\usepackage{epstopdf}
\usepackage[utf8]{inputenc}
\usepackage{algpseudocode}
\newcounter{qalgorithm}
\renewcommand{\theqalgorithm}{\arabic{qalgorithm}}
\begin{document}
\preprint{APS/123-QED}

\title{Evaluation-efficient quantum architecture search with ZX-calculus-based topological reuse 
}% Force line breaks with \\

\author{Chenlu Li}
%\email{lichenlu@imust.edu.cn}
 %\homepage{https://orcid.org/0000-0002-5420-7192}
\affiliation{School of Integrated Circuit, Nanjing University of Science and Technology, Nanjing, Jiangsu 210094, China}
\affiliation{School of Digital and Intelligent Industry, Inner Mongolia University of Science and Technology, Baotou, Inner Mongolia 014010, China}

\author{Hui Zeng}
\email{zenghui@njust.edu.cn}
 %\homepage{https://orcid.org/0000-0002-7657-6714}
\affiliation{School of Integrated Circuit, Nanjing University of Science and Technology, Nanjing, Jiangsu 210094, China}
 
\author{Dazhi Ding}
\email{dzding@njust.edu.cn}
\affiliation{School of Integrated Circuit, Nanjing University of Science and Technology, Nanjing, Jiangsu 210094, China}
\altaffiliation{Authors to whom correspondence should be addressed: zenghui@njust.edu.cn and dzding@njust.edu.cn}%Lines break automatically or can be forced with \\


\begin{abstract}
Variational quantum algorithm is a leading approach for quantum chemistry and many-body physics on noisy intermediate-scale quantum devices. The performance depends strongly on the structure of the parameterized quantum circuits. Quantum architecture search (QAS) can automate \textit{ansatz} design. However, it requires repeated training and evaluation of many candidate circuits, leading to high evaluation cost. In this work, we propose a noise-aware quantum architecture search framework based on ZX-calculus topological reuse (ZX-QAS). The framework encodes the search space with a ternary Gray-code mapping and integrates a noise-aware quantum neural network with a ZX-calculus topological reuse mechanism. The effectiveness of the framework is validated through ground-state energy estimation tasks and one-dimensional transverse-field Ising model tasks under noisy conditions. The results show that the ZX-QAS exhibits stable convergence and remarkably reduces the cost of expensive evaluations under noisy conditions.
\end{abstract}

\maketitle

\section{Introduction}
Variational quantum algorithm (VQA) is widely viewed as a practical route for near-term noisy intermediate-scale quantum (NISQ) devices\cite{ref1}\cite{ref2}\cite{REF5}. By using a classical computer to optimize the circuit parameters, VQA can solve problems within limited coherence times and reduce the need for deep quantum circuits. They have shown considerable potential in quantum chemistry simulation, combinatorial optimization, and quantum machine learning\cite{QML}\cite{Quantum-machine-learning}. The performance is determined not only by the classical optimizer, but also by the \textit{ansatz}\cite{REF3}\cite{REF4}. Deep quantum circuits with strong entanglement can offer higher expressibility, but they are more sensitive to decoherence, gate noise, and barren plateaus\cite{REF6}\cite{REF7}\cite{REF17}. In contrast, shallow hardware-efficient \textit{ansatz}es are often easier to train, although they may lack sufficient variational freedom. A great challenge is to balance expressibility, trainability, and hardware overhead in VQA\cite{REF10}.

Quantum architecture search (QAS) aims to automatically identify quantum circuits for a given problem\cite{REF8}\cite{REF9}. It turns \textit{ansatz} design into a discrete search problem\cite{REF11}\cite{REF12}\cite{REF13}. The QAS uses methods such as evolutionary algorithm (EA)\cite{EA}\cite{EA1}\cite{EA2}, reinforcement learning\cite{RL-QAS}, or differentiable search\cite{Differentiable1}\cite{Differentiable} to explore large architecture spaces under the constraints of the target Hamiltonian and hardware. The main bottleneck of these methods is the high evaluation cost. During the search, many candidate quantum circuits must be repeatedly trained and evaluated to explore their ground-truth performance. Moreover, some candidates have different gate sequences but similar topological structures. Re-evaluating these candidates requires substantial computational resources. Efficient low-cost evaluation is thus important for improving the scalability and practical utility of QAS.

Intrinsic noise in quantum devices is another important factor that limits the VQA performance. Noise can induce qubit flips and decoherence, which may degrade the algorithm or even make it fail. A practical QAS framework should be noise-aware. It is essential to explicitly account for noise during the search and select \textit{ansatz}es that remain robust under noisy conditions.

 In this context, we propose a noise-aware quantum architecture search framework based on ZX-calculus topological reuse (ZX-QAS). The framework uses a ternary Gray-code mapping to encode the search space and combines a noise-aware quantum neural network with a ZX-calculus-based topological reuse mechanism. It employs a variable-depth evolutionary search algorithm to reduce the cost of expensive circuit evaluations while maintaining circuit performance. The main contributions of this work are as follows.

First, we introduce an implicit mapping mechanism based on a ternary Gray code. The search space grows rapidly with the number of qubits, and explicit enumeration of all gate combinations can lead to out-of-memory (OOM) errors. Our mapping mechanism does not generate or store the large Cartesian product space. Instead, it decodes the rotation gates and sparse CNOT structure of each layer only when a candidate \textit{ansatz} is evaluated. More importantly, the ternary Gray-code ensures that a one-step mutation in the evolutionary algorithm corresponds to the change of only one local quantum gate. This preserves local continuity in the \textit{ansatz} and avoids abrupt structural jumps.

Second, we introduce a ZX-calculus-based topological reuse mechanism to reduce redundant evaluations. The mechanism parses each quantum circuit into a ZX graph and simplifies it using ZX-calculus rewrite rules. We further extract ZX structural features and define the structural similarity between candidate circuits based on these features. The resulting ZX structural similarity is used to identify candidate circuits that have low structural novelty relative to previously evaluated circuits. For highly similar candidates, the algorithm reuses the result of the nearest evaluated circuit with a probability that depends on the search progress. In this way, the ZX structural features serve as a heuristic prior for reducing redundant evaluation cost.

Third, we use a noise-aware parameter-sharing strategy to improve robustness under noisy conditions. We train several supernet parameter pools under the noise model. During the search stage, candidates inherit these shared parameters as initial values. This reduces the instability caused by random initialization and noise.

This article is organized as follows: Sec. II introduces recent advances in QAS and ZX-calculus. Sec. III details the ZX-QAS framework, encompassing ternary Gray-code mapping search-space design, the noise-aware quantum neural-network model, and a ZX-calculus-based topological reuse mechanism. Sect. IV presents experimental results and analysis. Sec. V gives the main conclusions.

\section{Related work}
In recent years, QAS has become an important approach for automated quantum circuit design. Instead of fixing the\textit{ ansatz} in advance, QAS treats the quantum circuit as a searchable object. It builds a feedback loop among the search space, the search strategy, and the evaluation mechanism. The key challenge has shifted from generating circuits automatically to finding high-quality circuits reliably with low evaluation cost.

This section illustrates recent advancements across three dimensions: the search space, search and evaluation strategy, and ZX-calculus.

\subsection{Search space}
The search space defines the candidate circuits that QAS can explore\cite{REF14}. It also shapes the optimization landscape explored by the search strategy. Early QAS studies usually construct the search space at the gate, layer, or block level. Gate-level spaces are the most flexible, but they also have the largest combinatorial size. Layer- and block-level spaces are more stable, but they may reduce expressibility. Regardless of how it is constructed, a practical search space should represent the quantum correlations required by the target problem, comply with the depth and two-qubit-gate limitations of NISQ hardware, and ensure that small changes in the encoding correspond to local changes in the quantum circuit.

Recent studies have shown that diversity control in the search space can mitigate structural degradation in QAS\cite{diversity-control}. Previous work suggests that the search space should describe not only the available gates, but also the training and evaluation conditions. Correspondingly, the circuit architectures and training settings can be included in the automated search\cite{AutoML}. Some studies further represent quantum circuits as graphs and predict candidate performance from graph features\cite{graph-features}. These studies indicate that circuit connectivity is an important factor for QAS efficiency.

Consequently, the search space should account for circuit topology, expressibility, trainability, and hardware overhead, rather than merely specify a set of available gates. However, existing discrete encodings do not always preserve structural locality between neighboring candidates and may require substantial memory to represent large search spaces.

\subsection{Search and evaluation strategy}
The performance of the QAS depends on the search strategy and evaluation strategy\cite{ref15}\cite{REF16}. The search strategy determines how the search algorithm explores a large search space. Evaluation provides the feedback that guides the search algorithm. Training each candidate circuit from scratch to estimate its ground-truth performance incurs a prohibitive computational cost. Therefore, the research key point has gradually shifted from improving search algorithms to reducing evaluation cost. Prevailing strategies encompass training-free proxies, weight-sharing supernet\cite{EA}, graph-based predictors\cite{GNN}, meta-learned generators\cite{Meta-qas}\cite{meta-train}, and reinforcement-learning policies\cite{RL}.

Strategies to enhance QAS efficiency generally fall into two paradigms. The first paradigm removes low-potential candidates before full optimization. This can be done using proxy metrics that require no training or only limited training\cite{train-free}\cite{train-free2}. Other approaches combine multiple proxies through a mixture-of-experts mechanism\cite{train-free1}, learn architecture priors with generative models\cite{meta-train1}, or use reinforcement learning to search for general quantum architectures\cite{general}. The second paradigm leverages prior knowledge during training to reduce redundant exploration. Representative approaches include graph-based neural predictors exploiting circuit topological features, as well as knowledge-driven QAS frameworks\cite{Knowledge-driven}. 

These strategies mainly use prior information to accelerate the evaluation of candidate circuits. However, later stages of evolutionary search often sample candidates near local structural clusters. Although many candidate circuits correspond to distinct encoded sequences, they exhibit highly similar circuit depths, entanglement configurations, and topological complexities. Re-evaluating these low-novelty candidates can cause substantial computational waste.

\subsection{ZX-calculus}
 ZX-calculus is a graphical language for representing and transforming quantum circuits. Unlike matrix representations, which act directly on exponentially large unitary matrices, ZX-calculus represents quantum processes as graphs made of spiders and edges. Z and X spiders describe phase operations in different bases, while Hadamard edges describe basis changes. By applying rewrite rules such as spider fusion, identity removal, color change, and local complementation, a ZX graph can be simplified without changing the quantum operation it represents\cite{ZX}. Since these rules act directly on graph structures, ZX-calculus is well suited for identifying redundant gates, analyzing commuting structures, and optimizing two-qubit gates.

ZX-calculus has recently been used to rewrite or mutate candidate circuits. For example, ZX-graph rewriting has been combined with reinforcement learning\cite{RF-ZX}. Other studies have designed ZX-inspired genetic mutation operators for quantum architecture search\cite{EPJ-ZX}\cite{zx-operator}. These works show that ZX graphs can reveal and remove local redundancy in quantum circuits. Previous studies have mainly used ZX-calculus for circuit simplification or architecture mutation. Its potential as a structural prior for managing the evaluation budget in QAS remains largely unexplored.

In summary, existing QAS methods still face three related challenges: the memory cost of representing large search spaces, the lack of structural locality in discrete encodings, and repeated evaluation of topologically similar circuits. Motivated by these limitations, we develop an implicit ternary Gray-code mapping and use ZX-calculus structural information to support topology-aware evaluation reuse. The detailed framework is presented in Sec. III. 
\begin{figure}[htbp]
        \centering
        \includegraphics[width=0.5\textwidth]{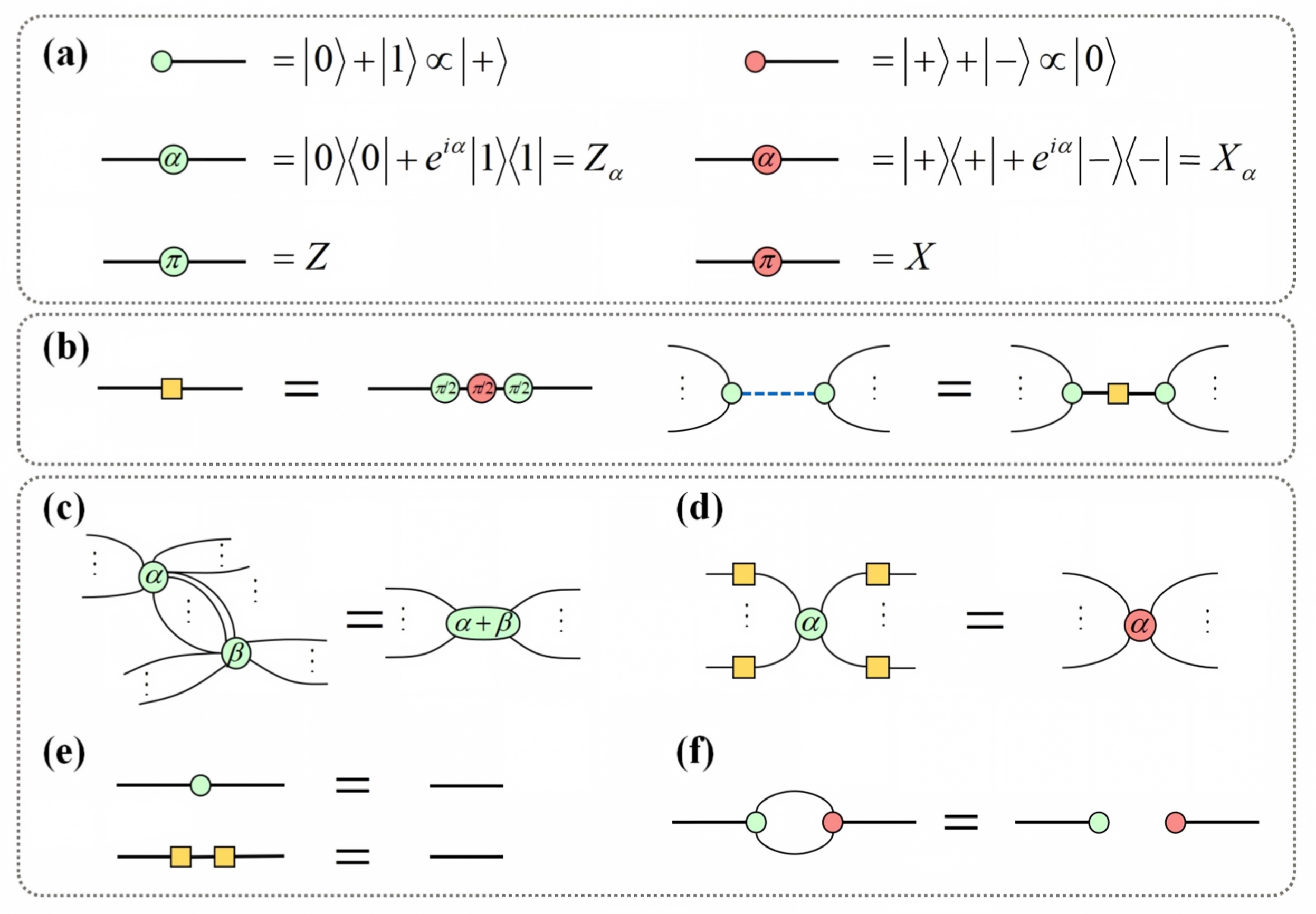}

                \caption{Basic graphical elements and rewrite rules of ZX-calculus. (a) State and phase-gate representations of Z and X spiders. Green nodes denote Z-basis spiders, and red nodes denote X-basis spiders. A single-input single-output spider with phase \(\pi\) corresponds to the Pauli-\(Z\) or Pauli-\(X\) operation. (b) Equivalent representations of the Hadamard gate and the Hadamard edge. (c) Spider-fusion rule. (d) Color-change rule. (e) Identity-removal rule. (f) Antipode rule. ZX-calculus contains other rewrite rules, which are not listed here; see Ref. \cite{ZX} for details.}
        \label{FIG.1}
\end{figure}

\section{Methodology}
As quantum circuits grow in size, the search space grows exponentially. Explicitly storing all candidate \textit{ansatzes} can lead to OOM errors. During search and evaluation, many candidate \textit{ansatze}s have similar topological structures, but they still require repeated parameter training and gradient evaluation. This results in substantial computational waste.

We propose a noise-aware quantum architecture search framework based on ZX-calculus topological reuse. Based on variational-quantum-circuit design, this framework introduces an implicit mapping mechanism based on a ternary Gray code. In the on-demand implementation, the memory footprint is determined by the current candidate encoding and the sparse CNOT lookup table, rather than by the full number of possible layer configurations. It also preserves topological locality and stabilizes the evolutionary search. The ZX-QAS framework uses a noise-aware quantum architecture training model and obtains noise-robust \textit{ansatz} parameters through parameter sharing. In the search stage, we introduce a ZX-calculus-based topological reuse mechanism to reduce repeated evaluations of low-novelty topologies. 

\begin{figure*}
        \centering
        \includegraphics[width=1\linewidth]{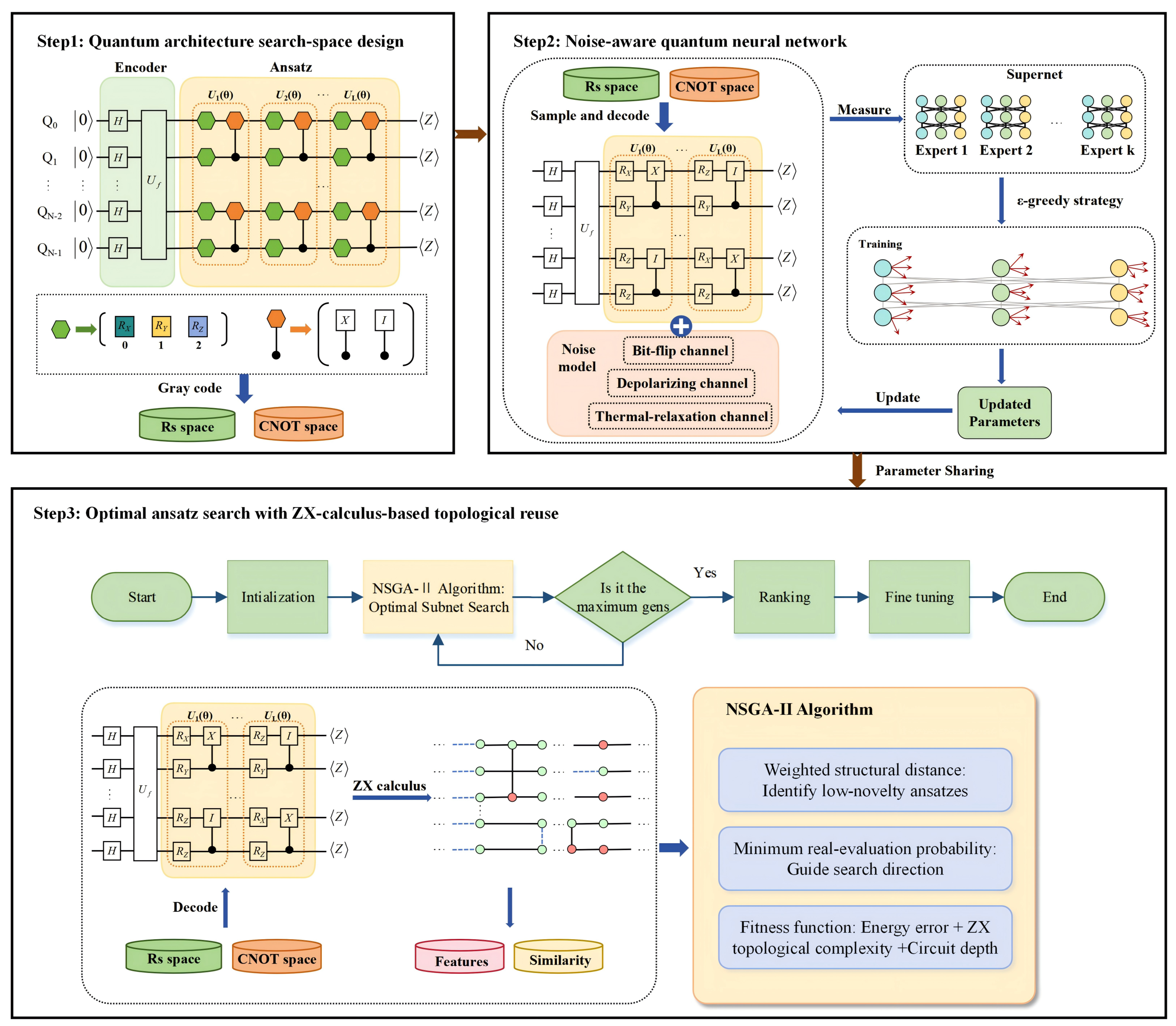}

                \caption{Workflow of the ZX-QAS framework. In step 1, single-qubit rotation gates are mapped to the \(R_s\) space using a ternary Gray-code mapping, and entangling gates are represented by a sparse CNOT lookup table. In step 2, a sampled \textit{ansatz} is decoded and combined with the noise model, and multiple supernet experts are trained. The shared parameters are updated through an \(\epsilon\)-greedy strategy. In step 3, the search inherits the noise-aware supernet parameters. During evolutionary search, each candidate \textit{ansatz} is converted into a ZX graph for topological feature extraction. A weighted structural distance, a minimum real-evaluation probability, and a fitness function are used to implement ZX-calculus-based topological reuse and optimal \textit{ansatz} search.}
        \label{FIG.2}
\end{figure*}

As show in Figure \ref{FIG.2} , the framework consists of three main components: quantum architecture search-space design, a noise-aware quantum neural network, and QAS with ZX-calculus-based topological reuse.

\subsection{Ternary Gray-code mapping for the search space}
We construct a structured quantum architecture search space. This search space consists of a single-qubit rotation-gate subspace and an entangling-gate subspace.

The rotation subspace uses the universal single-qubit rotation gates $\{R_x, R_y, R_z\}$, representing rotations around the $x$, $y$ , and $z$ axes of the Bloch sphere, respectively. For an  \(n\)-qubit  \textit{ansatz}, each qubit is allowed to choose its rotation axis independently. The set of all possible single-layer rotation-gate combinations is given by:
\begin{equation}
R_{\text{space}} = \{ R_x, R_y, R_z \}^{n}.
\label{Eq.1}
\end{equation}

The size of the rotation subspace is $|R_{\text{space}}|=3^n$. 

The entangling subspace uses CNOT gates. To avoid over-parameterization and reduce the risk of barren plateaus, each layer contains at most one CNOT gate. 
\begin{equation}
  \text{CNOT}_{\text{space}} = \{\emptyset\}\cup\left\{e\mid e\in E_{\rm logic}\right\} . 
  \label{Eq.2}
\end{equation}

Here \(E_{\rm logic}\) is the set of all possible two-qubit pairs, with \(|E_{\rm logic}|=n(n-1)/2\). Thus, the size of the entangling subspace is 
\begin{equation}
  |\text{CNOT}_{\text{space}}|=1+|E_{\rm logic}|. 
  \label{Eq.3}
\end{equation}

Explicitly enumerating and storing the Cartesian product of \(R_{\text{space}}\) and \(\text{CNOT}_{\text{space}}\) can lead to OOM errors. Standard integer encoding lacks structural locality. When the index is increased by one, several digits may change at the same time, which corresponds to simultaneous changes of multiple quantum gates.

To address these limitations, we introduce a hybrid implicit mapping strategy based on the different structures of the two subspaces. The rotation subspace is encoded by a ternary Gray-code \(G_3(n)\). It is defined recursively as follows. For one dimension, \(G_3(1)=(0,1,2)\). For any \(n>1\), \(G_3(n)\) is constructed by reflecting and concatenating the \((n-1)\)-dimensional sequence, and then appending the symbols \(\{0,1,2\}\) in order. This construction guarantees that the Hamming distance between any two adjacent \(n\)-dimensional codes is always one:
\begin{equation}
 \mathcal{D}_H(\mathbf{g}_u,\mathbf{g}_{u+1})\equiv 1.  
  \label{Eq.4}
\end{equation}

The CNOT subspace is indexed by a sequential lookup table. Each layer is represented by an integer \(u\in[0,|R_\text{space}| \times |\text{CNOT}_\text{space}|-1]\). The quantum gates are decoded only when the candidate is evaluated:

\begin{equation}
\begin{aligned}
\mathrm{Rot}_{\mathrm{Idx}}(u)
&=
\left\lfloor
\frac{u}{\left|\mathrm{CNOT}_{\mathrm{space}}\right|}
\right\rfloor,\\
\mathrm{CNOT}_{\mathrm{Idx}}(u)
&=
u \bmod
\left|\mathrm{CNOT}_{\mathrm{space}}\right|.
\end{aligned}
\label{Eq.5}
\end{equation}
\begin{equation}
\Phi(u)
=
\left(
G_{3}(n)\left[\mathrm{Rot}_{\mathrm{Idx}}(u)\right],
\mathrm{CNOT}_{\mathrm{space}}
\left[\mathrm{CNOT}_{\mathrm{Idx}}(u)\right]
\right),
\label{Eq.6}
\end{equation}
where $\Phi(u)\in
R_{\mathrm{space}}\times\mathrm{CNOT}_{\mathrm{space}}$. Eq.~\eqref{Eq.5} decomposes the integer index $u$ into a rotation-subspace index and an entangling-subspace index. For a fixed $\mathrm{Rot}_{\mathrm{Idx}}(u)$, increasing $u$ enumerates all $\lvert\mathrm{CNOT}_{\mathrm{space}}\rvert$ entangling configurations, with $\mathrm{CNOT}_{\mathrm{Idx}}(u)
=0,1,\ldots,\lvert\mathrm{CNOT}_{\mathrm{space}}\rvert-1$. After the last entangling configuration is reached, the next value of $u$ resets $\mathrm{CNOT}_{\mathrm{Idx}}(u)$ to zero and increments $\mathrm{Rot}_{\mathrm{Idx}}(u)$, thereby moving to the next rotation configuration. Eq. ~\eqref{Eq.6} maps $u$ to an ordered pair in $R_{\mathrm{space}}\times\mathrm{CNOT}_{\mathrm{space}}$. Here, $G_3(n)[\mathrm{Rot}_{\mathrm{Idx}}(u)]$ denotes the rotation-gate configuration obtained from the $n$-dimensional ternary Gray-code sequence, while $\mathrm{CNOT}_{\mathrm{space}} [\mathrm{CNOT}_{\mathrm{Idx}}(u)]$ selects the corresponding entangling configuration from the CNOT lookup table. These two entries specify the gate configuration of a single circuit layer.

The reflection property of the Gray code guarantees that, within the rotation subspace, two adjacent indices \(u\) and \(u+1\) differ only in one single-qubit rotation gate. This preserves the locality of genetic operations in the subsequent evolutionary search. A small mutation of the genetic code therefore produces only a local gate change, rather than an abrupt structural change of the quantum circuit.

\subsection{Noise-aware quantum neural networks}
To improve the efficiency and reliability of candidate evaluation, we adopt a supernet pretraining module based on the single-path one-shot (SPOS) strategy. Following Li et al. \cite{ours}, the supernet is trained under a noise model, providing shared parameters that are adapted to the noisy environment and used to initialize candidate circuits.

The noise model is implemented by using MindQuantum framework and it encompasses three noise channels that characterize gate operations in real quantum hardware: 

(1) Bit-flip Channel: The bit-flip channel is modeled by applying a Pauli-$X$ operator with probability $p_{BF}$ following each quantum gate operation. The output quantum state is described by the density matrix:  

\begin{equation}
 \varepsilon_{BF}(\rho) = (1-p_{BF})I\rho I + p_{BF}X\rho X.
\label{Eq.7}    
\end{equation}

where \(\rho\) is the density matrix of the input quantum state.

(2) Depolarizing Channel: The depolarizing channel describes a process that replaces the quantum system with a completely mixed state with a certain probability $p$. For a quantum gate acting on $n$ qubits, the quantum operation of the depolarizing channel can be uniformly expressed as:
\begin{equation}
    \varepsilon_{DF}(\rho) = (1-p)\rho + \frac{p}{4^n - 1} \sum_{k=1}^{4^n-1} P_k \rho P_k^\dagger.
    \label{Eq.8}
\end{equation}

where $\rho$ is the density matrix of the system, and $P_k$ represents the non-trivial tensor product operations in the $n$-qubit Pauli group (excluding the identity matrix $I^{\otimes n}$).

Specifically, for single-qubit rotation gates (e.g., $R_X$, $R_Y$, and $R_Z$), the above formula reduces to $n=1$, indicating the occurrence of three basic Pauli errors ($X$, $Y$, and $Z$) with equal probability ($p_1/3$). 

However, on real superconducting quantum computers in the noisy intermediate-scale quantum (NISQ) era, the execution of two-qubit entangling gates (such as the CNOT gate) not only incurs higher error rates but is also highly susceptible to severe crosstalk and correlated errors. Therefore, to more accurately fit the decoherence characteristics of physical hardware, we extend the depolarizing channel acting on CNOT gates to the complete two-qubit space (i.e. $n=2$). In this case, the error model covers $4^2 - 1 = 15$ combinations of two-qubit Pauli operators (such as $I \otimes X$, $X \otimes Y$, $Z \otimes Z$, etc.), which occur with an equal probability of $p_2/15$.

(3) Thermal Relaxation Channel: The thermal relaxation channel describes energy relaxation (characterized by time $T_{1}$) and phase relaxation/decoherence (characterized by time $T_{2}$) during a quantum gate operation of duration $T_{g}$. After thermal relaxation, the state is described by:
\begin{equation}
\varepsilon_{TR}(\rho) = \text{tr}_1\left[\Lambda\left(\rho^T \otimes I\right)\right], \quad \Lambda = \begin{pmatrix}
\varepsilon_{T_1} & 0 & 0 & \varepsilon_{T_2} \\
0 & 1-\varepsilon_{T_1} & 0 & 0 \\
0 & 0 & 0 & 0 \\
\varepsilon_{T_2} & 0 & 0 & 1
\end{pmatrix}.
\label{Eq.9}
\end{equation}

where \(\varepsilon_{T_1} = e^{-T_g/T_1}\)  and \(\quad \varepsilon_{T_2} = e^{-T_g/T_2}\) denote the probabilities of the qubit remaining unaffected by energy relaxation and decoherence, respectively, within the interval $T_{g}$. \(\quad \Lambda\)  denotes the Choi matrix, representing the complete probability distribution of all possible outcomes of the thermal relaxation channel on the quantum state within the time interval $T_{g}$.

To reduce the cost of candidate evaluation, we construct $K$ parameter-sharing supernet experts, each of which maintains an independent parameter pool. During SPOS pretraining, an ansatz architecture $\mathcal{A}$ is
randomly sampled from the integer-encoded search space. For each expert $k\in\{1,\ldots,K\}$, the sampled circuit is parameterized with $\boldsymbol{\theta}_k$. Let $\rho(\mathcal{A},\boldsymbol{\theta}_k)$ denote the output density matrix obtained under the specified noise model. The corresponding energy expectation is
\begin{equation}
E_k=\operatorname{Tr}\left[\rho(\mathcal{A},\boldsymbol{\theta}_k)H\right].
\label{Eq.10}
\end{equation}

To guide the variational optimization and reduce the risk of barren plateaus, we first select the expert with the lowest current energy, \(k^*=\arg\min_k E_k\). An \(\epsilon\)-greedy strategy is then used to update the parameter pools. With probability \(1-\epsilon\), the best-performing expert \(\tilde{k}=k^*\) is exploited, while with probability \(\epsilon\), another expert is randomly selected for exploration. The shared parameters of the selected expert \(\tilde{k}\) are updated by gradient descent:

\begin{equation}
\boldsymbol{\theta}_{\tilde{k}}^{(t+1)} = \boldsymbol{\theta}_{\tilde{k}}^{(t)} - \eta \nabla_{\boldsymbol{\theta}} E_{\tilde{k}}.
\label{Eq.11}
\end{equation}

This hybrid \(\epsilon\)-greedy strategy helps stabilize the optimization process. The pretrained weights and parameters are then inherited by the subsequent evolutionary search as the initial parameter \(\boldsymbol{\theta}_0\).

\subsection{Quantum architecture search with ZX-calculus-based topological reuse}

After constructing the implicit search space and pretraining the noise-aware supernet, we perform a variable-depth evolutionary search using the evolutionary operators implemented in nondominated sorting genetic algorithm II (NSGA-II). Candidate circuits are ranked according to a scalar fitness function. To reduce the computational redundancy caused by repeated evaluations of candidates with similar topology, we introduce a ZX-calculus-based topological reuse mechanism into the fitness-evaluation stage, as shown in Algorithm\ref{alg:zx_qas}. The algorithm proceeds as follows.

\textit{Step 1: Population initialization. }

We initialize a parent population \(P_0\) of size \(N_{\text{pop}}\). Each individual \(\mathcal{A}_i\) is represented by a variable-length one-dimensional array of scalar integers. Each integer is sampled from \([0, |R_{\text{space}}|\times |\text{CNOT}_{\text{space}}|-1]\) and encodes one computational layer of the quantum circuit.

\textit{Step 2: ZX-calculus-based topological reuse.}

Before gradient optimization, the decoded candidate \textit{ansatz} \(\mathcal{A}_i\) is first parsed into a ZX graph \(G_i=(V,E)\). We then apply ZX-calculus rewrite rules to reduce it to an abstract graph \(G'_i\), and record the ZX structural-feature signature \(\sigma(\mathcal{A}_i)\). This signature includes the number of nodes, node types, node phases, the qubit/row index of each node, edge connections, and edge types. 

If the exact architecture encoding has appeared before, its cached fitness is reused directly. In addition, if the reduced ZX structural signature \(\sigma(\mathcal{A}_i)\) matches that of an evaluated candidate, the two candidates are treated as having the same reduced ZX structural pattern under the signature used in this work, and the cached fitness is reused.

If no topological match is found, we further compute the ZX similarity features of the candidate. This step is heuristic. It does not assume exact circuit equivalence. Instead, it uses structural similarity as a prior for reducing redundant evaluations. These features include the circuit depth, the number of spiders after ZX reduction, the number of CNOT gates, the total numbers of the three types of single-qubit rotation gates, and the distribution of entangling edges. Let \(\mathbf{s}(\mathcal{A}_i)\) denote the ZX similarity feature vector of \(\mathcal{A}_i\), and let \(\mathcal{B}\) denote the set of architectures that have been evaluated. We find the nearest evaluated candidate \(\mathcal{A}_{j^*}\) in \(\mathcal{B}\), defined as
\begin{equation}
j^*=\arg\min_{j\in\mathcal{B}} D\!\left(\mathbf{s}(\mathcal{A}_i),\mathbf{s}(\mathcal{A}_j)\right).
\label{Eq.12}
\end{equation}

Here, \(D(\cdot,\cdot)\) is a weighted normalized structural distance. The number of spiders after ZX reduction and the distribution of entangling edges are assigned relatively large weights, because they directly reflect the topology and entanglement structure of the ZX graph. The circuit depth is assigned an intermediate weight, as it affects both expressibility and noise accumulation. The number of CNOT gates and the distribution of single-qubit rotation gates are used as auxiliary features to distinguish local gate structures. When \(D(\mathbf{s}(\mathcal{A}_i),\mathbf{s}(\mathcal{A}_{j^*}))<\tau\), the candidate \textit{ansatz} is regarded as a low-novelty structure.

For a low-novelty candidate, the algorithm does not reuse a historical result directly. Instead, the reuse probability increases gradually as the evolutionary search proceeds, while a minimum probability of real evaluation is always retained:
\begin{equation}
p_{\rm real}(t)=\max\left(p_{\min},1-\frac{t}{T}\right).
\label{Eq.13}
\end{equation}

Here, \(t\) denotes the current search progress, \(T\) denotes the total search budget, and \(p_{\min}\) is the minimum probability of real evaluation. At the early stage of the search, \(p_{\rm real}(t)\) is relatively large, so the algorithm mainly relies on real evaluations to build a reliable cache. At later stages, \(p_{\rm real}(t)\)decreases, and the algorithm becomes more likely to reuse historical evaluation results to reduce repeated evaluation cost.

If the candidate \textit{ansatz} \(\mathcal{A}_i\) is treated as a low-novelty architecture and reuse is triggered, its fitness is computed from the fitness of the nearest evaluated \textit{ansatz} \(\mathcal{A}_{j^*}\) with an additional redundancy penalty:
\begin{equation}
F_{\rm reuse}(\mathcal{A}_i)=F(\mathcal{A}_{j^*})-R_i(t),
\label{Eq.14}
\end{equation}
\begin{equation}
R_i(t)=\gamma_{\max}\frac{t}{T}\left[1-\frac{D\!\left(\mathbf{s}(\mathcal{A}_i),\mathbf{s}(\mathcal{A}_{j^*})\right)}{\tau}\right].
\label{Eq.15}
\end{equation}

Here, \(\gamma_{\max}\) is the maximum reuse-penalty coefficient. This penalty is introduced only when a low-novelty candidate is actually reused. Its magnitude depends on both the search progress and the structural distance between the two candidates. A smaller distance indicates that the candidate is closer to an evaluated structure, and therefore gives a larger redundancy penalty. As the search approaches later stages, repeated structures are also penalized more strongly. If the candidate is not identified as low-novelty, or if reuse is not triggered, it is still evaluated by the normal evaluation procedure.

When a candidate \textit{ansatz} needs to be evaluated, the algorithm selects the parameter set with the lowest energy from the noise-aware supernet expert pools as the initial point. It then performs local fine tuning with L-BFGS-B and obtains the optimized parameters \({\theta}_i^*\). The evaluated energy of this \textit{ansatz} is \(E_i^*=E(\mathcal{A}_i,\boldsymbol{\theta}_i^*)\). With the reference ground-state energy \(E_{\rm ref}\), the energy error is defined as
\begin{equation}
\Delta E_i = |E_i^* - E_{\rm ref}|.
\label{Eq.16}
\end{equation}

We include the energy error, the ZX topological complexity, and the circuit depth in a scalar fitness function:
\begin{equation}
F_{\rm eval}(\mathcal{A}_i)=-\left[\Delta E_i+\alpha C_{\rm ZX}(\mathcal{A}_i)+\beta L(\mathcal{A}_i)\right].
\label{Eq.17}
\end{equation}

Here, \(C_{\rm ZX}(\mathcal{A}_i)\) is the number of spiders that remain after the candidate circuit is simplified by ZX-calculus. It measures the residual topological complexity of the reduced graph. \(L(\mathcal{A}_i)\) is the number of circuit layers used in the evaluation. It approximately accounts for circuit execution time, gate-noise accumulation, and decoherence risk. The constants \(\alpha\) and \(\beta\) are weighting coefficients.

The complete fitness rule is therefore
\begin{align}
F(\mathcal{A}_i)
&=
\begin{cases}
F(\mathcal{A}_j), & \mathcal{A}_i=\mathcal{A}_j
\ \text{or}\ 
\sigma_{\rm ZX}(\mathcal{A}_i)=\sigma_{\rm ZX}(\mathcal{A}_j),\\
F_{\rm reuse}(\mathcal{A}_i), & \mathcal{A}_i \in \Omega_{\rm reuse},\\
F_{\rm eval}(\mathcal{A}_i), & \text{otherwise}.
\end{cases}
\label{Eq.18}
\end{align}
where\begin{equation}
\Omega_{\rm reuse}
=
\left\{
\mathcal{A}_i\mid
D\!\left(\mathbf{s}(\mathcal{A}_i),
\mathbf{s}(\mathcal{A}_{j^*})\right)\le \tau
\ \text{and soft reuse is triggered}
\right\}.
\label{Eq.19}
\end{equation}

\textit{Step 3: Fitness-based ranking and selection.}

After fitness evaluation, the current population is ranked and selected according to the fitness.

\textit{Step 4: Selection, crossover, and mutation.}

Parent individuals are generated by tournament selection. Crossover and mutation are then applied to the integer-encoded chromosomes. Because the rotation subspace is encoded by a ternary Gray-code mapping, a local change in the integer encoding usually corresponds to a local gate change in the physical circuit. This helps preserve the structural correlation between candidate circuits.

\textit{Step 5: Elitism and population update.}

The parent population \(P_t\) and the offspring population \(Q_t\) are combined. The top \(N_{\rm pop}\) individuals are then selected according to the scalar fitness to form the next population \(P_{t+1}\). This process is repeated until the maximum number of generations is reached. The algorithm finally returns the best candidate architecture found during the search.

\refstepcounter{qalgorithm}
\par\medskip
\noindent\textbf{Algorithm \theqalgorithm.
QAS with ZX-calculus-based topological reuse}
\label{alg:zx_qas}

\begin{algorithmic}[1]
\Require Target Hamiltonian $H$, maximum generations $G_{\max}$, population size $N_{\rm pop}$, pretrained supernet parameters $\{\boldsymbol{\theta}_k\}_{k=1}^{K}$
\Ensure Best quantum architectures under the scalar fitness

\State Initialize population $P_0$ with integer-encoded circuits
\State Initialize the exact architecture cache, strict ZX signature cache, and ZX feature bank

\For{$t = 0$ to $G_{\max}-1$}
    \For{each architecture $\mathcal{A}_i \in P_t$}
        \State Decode $\mathcal{A}_i$ into the actual circuit

        \If{$\mathcal{A}_i$ exists in the exact architecture cache}
            \State \Return cached fitness $F(\mathcal{A}_i)$
        \EndIf

        \State Parse $\mathcal{A}_i$ into a ZX graph $G_{\rm ZX}$
        \State Reduce $G_{\rm ZX}$ to $G'_{\rm ZX}$
        \State Compute the strict signature $\sigma(\mathcal{A}_i)$ and spider count $C_{\rm ZX}(\mathcal{A}_i)$

        \If{$\sigma(\mathcal{A}_i)$ exists in the strict ZX signature cache}
            \State \Return cached fitness $F(\mathcal{A}_j)$
        \EndIf

        \State Extract the ZX feature vector $\mathbf{s}(\mathcal{A}_i)$
        \State Find the nearest evaluated architecture $\mathcal{A}_{j^\ast}$
        \State Compute $p_{\rm real}(t)=\max(p_{\min},1-t/T)$

        \If{$D(\mathbf{s}(\mathcal{A}_i),\mathbf{s}(\mathcal{A}_{j^\ast})) \le \tau$ and soft reuse is triggered}
            \State Compute $R_i(t)$
            \State $F(\mathcal{A}_i) \gets F(\mathcal{A}_{j^\ast}) - R_i(t)$
            \State \Return $F(\mathcal{A}_i)$
        \Else
            \State $\boldsymbol{\theta}_0 \gets$ best pretrained supernet parameters
            \State $\boldsymbol{\theta}_i^\ast \gets$ L-BFGS-B fine tuning under the noise model
            \State $E_i^\ast \gets E(H,\mathcal{A}_i,\boldsymbol{\theta}_i^\ast)$
            \State $\Delta E_i \gets |E_i^\ast - E_{\rm ref}|$
            \State $F(\mathcal{A}_i) \gets -\left[\Delta E_i+\alpha C_{\rm ZX}(\mathcal{A}_i)+\beta L(\mathcal{A}_i)\right]$
            \State Store $\mathcal{A}_i$, $\sigma_{\rm ZX}(\mathcal{A}_i)$, $\mathbf{s}(\mathcal{A}_i)$, $E_i^\ast$, $\Delta E_i$, and $F(\mathcal{A}_i)$
        \EndIf
    \EndFor

    \State Generate offspring $Q_t$ through selection, crossover, and mutation
    \State Select the top $N_{\rm pop}$ individuals according to the scalar fitness to form $P_{t+1}$
\EndFor

\State \Return best architectures found during the search
\end{algorithmic}
\par\medskip

\section{Evaluation}
We evaluate the feasibility and performance of ZX-QAS using ground-state energy estimation and the one-dimensional transverse-field Ising model (1D TFIM). Under noisy conditions, we compare different QAS frameworks in terms of error convergence, the relation between ZX topological complexity and energy error, and the reuse rate of candidate \textit{ansatz}es. We also perform ablation experiments to verify the effectiveness of the ternary Gray-code implicit mapping mechanism.

\subsection{Experimental Setup}
To evaluate the performance of ZX-QAS framework under realistic NISQ conditions, we establish a baseline hardware noise configuration aligned with the average performance metrics of current mainstream superconducting quantum processors. 

In our fundamental simulation experiments, the baseline noise parameters ( $\lambda=1.0$ ) are configured as follows: the single-qubit and two-qubit depolarizing probabilities are set to $p_1 = 0.001$ and $p_2 = 0.004$, respectively. The measurement bit-flip error rate is $p_{BF} = 0.05$. For the thermal relaxation channel, the relaxation time is $T_1 = 100,000$ ns, the dephasing time is $T_2 = 50,000$ ns, and the average single-qubit and two-qubit gate execution duration is $T_{g_1} = 30$ ns and $T_{g_2} = 80$ ns, respectively. 

\subsection{Ground-state energy estimation}
To evaluate the physical expressibility of the proposed ZX-QAS framework, we perform variational quantum eigensolver (VQE) experiments for the ground-state energies of $H_2$, $H_4$, and $LiH$ under noisy conditions.

For each molecule, the Hamiltonian is generated from the atomic species and relative atomic coordinates using OpenFermion and OpenFermion-PySCF with the STO-3G basis. The \(H_2\) molecule is set with an H-H bond length of 0.7414 \(\AA\). The \(H_4\) molecule is taken as a linear hydrogen chain with a nearest-neighbor spacing of 1.0  \(\AA\). The LiH molecule is set with a Li-H bond length of 1.546  \(\AA\). The \(H_2\) and \(H_4\) Hamiltonians require 4 and 8 qubits, respectively. The full LiH Hamiltonian requires 12 qubits. For LiH, we freeze the Li \(1s\) core orbital and keep orbitals 1, 2, and 3 as the active space\cite{LiH-RL}. This gives a 6-qubit active-space Hamiltonian with two active electrons, which is mapped by the Jordan-Wigner transformation. The reference ground-state energies are obtained from FCI or exact diagonalization of the corresponding Hamiltonians. The number of supernet experts is set to $K=5$, and the circuit depth is set as $l_{\min}=3$ and $l_{\max}=10$. For the ZX topological reuse mechanism, the similarity threshold is set to $\tau=0.23$, the maximum redundancy-penalty coefficient to $\gamma_{\max}=0.035$, and the minimum real-evaluation probability to $p_{\min}=0.12$.

\begin{figure*}
    \centering
    {\includegraphics[width=1\linewidth]{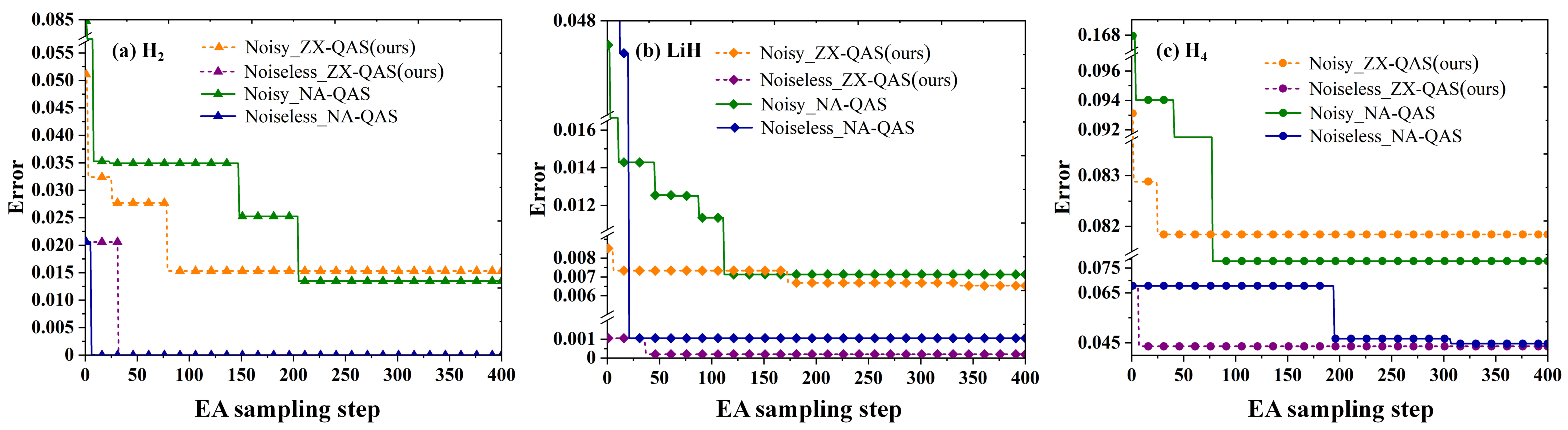}}
 
    \caption{Energy-error convergence curves. Panels (a), (b), and (c) show the results for $H_2$, $LiH$, and $H_4$, respectively. The horizontal axis denotes the evaluation step, and the vertical axis denotes the best-so-far energy error. The orange and purple curves show the results of the proposed ZX-QAS under noisy and noiseless conditions, respectively. The green and blue curves show the results of NA-QAS without ternary Gray-code implicit mapping and ZX topological reuse under noisy and noiseless conditions, respectively.}
    \label{FIG.3}
\end{figure*}

Figure \ref{FIG.3} compares the error convergence of ZX-QAS and NA-QAS for three molecular ground-state energy estimation tasks under noisy and noiseless conditions. Overall, ZX-QAS reaches errors comparable to, or lower than, those of the baseline in most cases with fewer evaluation steps. This indicates that ZX topological reuse does not destroy the main convergence direction of the evolutionary search.

For $H_2$ (Fig. \ref{FIG.3}(a)), under noisy conditions, the error of ZX-QAS decreases rapidly from about 0.051 hartree and stabilizes near 0.015 hartree after about 80 evaluations. In comparison, NA-QAS reaches a slightly lower error at the later stage, but its main improvement occurs later in the search. Under noiseless conditions, both methods converge to nearly zero error. This suggests that the \textit{ansatz} search space is expressive enough for this small system, and that noise is the main factor limiting the final accuracy.

For $LiH$ (Fig. \ref{FIG.3}(b)), the advantage of ZX-QAS is more pronounced. Under noisy conditions, ZX-QAS rapidly reaches an error of about 0.007 hartree at the early stage and then decreases steadily toward convergence. NA-QAS shows a longer staircase-like descent. Under noiseless conditions, ZX-QAS reaches an error close to zero and outperforms NA-QAS. This indicates that, for a more complex molecular system, ZX topological reuse can reduce repeated evaluations of low-novelty structures while preserving the search ability in promising structural regions.

For $H_4$ (Fig. \ref{FIG.3}(c)), under noisy conditions, ZX-QAS reaches a final error of about 0.082 hartree, slightly higher than the value of about 0.078 hartree obtained by NA-QAS. This suggests that the ZX topological reuse mechanism may introduce a small loss in accuracy for the larger hydrogen chain. Under noiseless conditions, ZX-QAS reaches an error of about 0.044 hartree at an early stage and remains stable, which is slightly better than the final error of about 0.046 hartree obtained by noiseless NA-QAS. This shows that the search mechanism of ZX-QAS does not reduce the circuit expressibility. The difference under noisy conditions is more likely related to noise accumulation and uncertainty in structural-similarity estimation for the more complex system.

These results show that the main contribution of ZX-QAS is not to minimize the energy error at any cost. Rather, it reduces repeated evaluations through structure-aware reuse while maintaining stable convergence within an acceptable accuracy loss. For $H_2$ and $LiH$, ZX-QAS is competitive in both convergence speed and final error. For $H_4$, although the error is slightly higher than that of the baseline under noisy conditions, the result still demonstrates the feasibility of a cost-accuracy tradeoff in a more complex molecular system.

\begin{figure*}
    \centering
      {\includegraphics[width=1\linewidth]{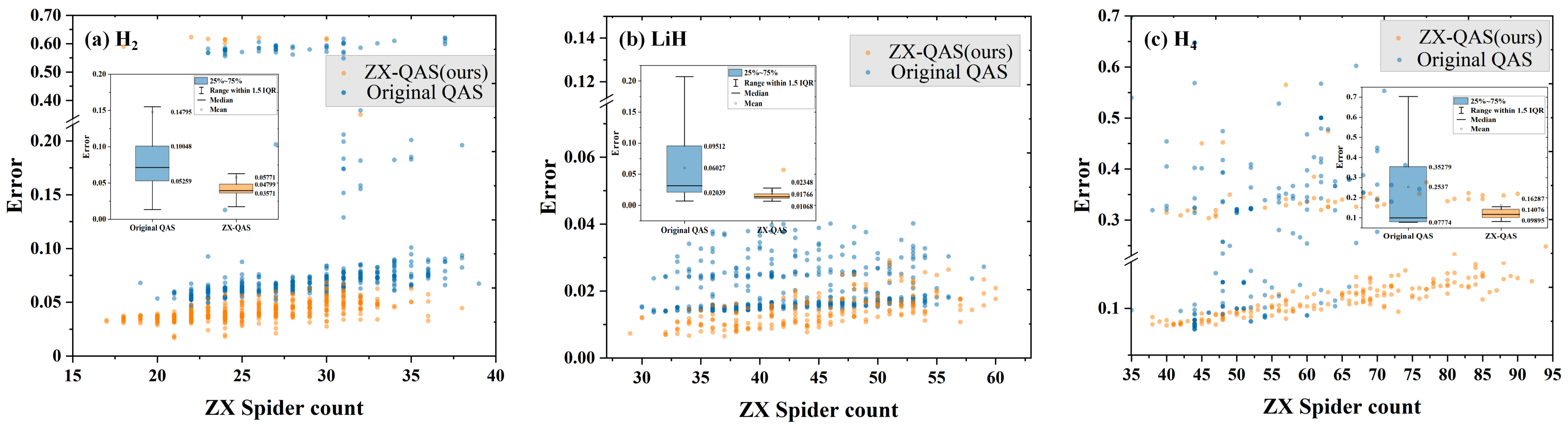}}

    \caption{Relation between ZX topological complexity and molecular ground-state energy error. Panels (a), (b), and (c) show the results for $H_2$, $LiH$, and $H_4$, respectively. The main plots show the distribution of candidate-\textit{ansatz} errors as a function of the spider count after ZX simplification. The insets show box plots of the error distributions for ZX-QAS and Original QAS. Orange points denote the proposed ZX-QAS, while blue points denote Original QAS without ternary Gray-code implicit mapping and ZX topological reuse.}
    \label{FIG.4}
\end{figure*}

Figure \ref{FIG.4} shows the relation between the topological complexity after ZX simplification and the energy error. Overall, the points of ZX-QAS are more concentrated in the low-error region, whereas Original QAS shows a broader error distribution and more high-error outliers. This indicates that the ZX topological reuse mechanism does not merely reduce the number of evaluated candidates. Instead, it guides the search toward structural regions with more stable and lower-error performance.

For $H_2$ (Fig. \ref{FIG.4}(a)), the ZX-QAS points are mainly concentrated in the error range of 0.03-0.06 hartree. In contrast, Original QAS contains many high-error candidates above 0.1 hartree, with some errors approaching 0.6 hartree. The box plot in the inset further shows that both the median error and the interquartile range of ZX-QAS are lower than those of Original QAS. This suggests that, for a small molecular system, the ZX-based structural constraint can effectively suppress low-quality candidate circuits.

For $LiH$ (Fig. \ref{FIG.4}(b)),  the ZX-QAS points are concentrated near an error of 0.01 hartree. The box plot shows a narrower error distribution and a lower median than Original QAS. This indicates that ZX-QAS can maintain stable search quality in a molecular system with more complex interactions. By contrast, Original QAS has a broader distribution, which reflects larger fluctuations in candidate-circuit performance.

 For $H_4$ (Fig. \ref{FIG.4}(c)), the task complexity increases significantly. Original QAS produces many high-error candidates, with errors exceeding 0.6 hartree in some cases. This shows that, in a larger hydrogen chain, a search without structural reuse and topological constraints is more likely to sample low-quality architectures. Although the errors of ZX-QAS also increase with the spider count, their overall distribution is more concentrated, and high-error outliers are clearly reduced. In the inset, ZX-QAS has a narrower box, and both its median and mean errors are lower than those of Original QAS. This suggests that ZX-QAS improves the stability of candidate-circuit quality in more complex systems.

From the viewpoint of topological complexity, the three experiments show that the energy error is not a monotonic function of the spider count. A smaller number of ZX spiders usually corresponds to a simpler circuit with less noise accumulation, but it may also imply insufficient expressibility. A larger number of spiders may provide stronger expressibility, but it can also introduce more noise and make the optimization harder. Therefore, this work does not use the spider count as the only criterion for circuit quality. Instead, it is used as a structural regularization term in the fitness function and is combined with ZX similarity features for topological reuse.

Figure \ref{FIG.4} shows that the topological features after ZX simplification are statistically correlated with candidate-circuit performance. Therefore, they can serve as structural priors in the evaluation stage. Compared with Original QAS, ZX-QAS compresses the error distribution, reduces high-error candidate circuits, and improves the stability of the search results. These results indicate that ZX-QAS not only reduces the evaluation cost, but also improves the quality of candidate-circuit screening to some extent.

\begin{figure*}
    \centering
    \includegraphics[width=1\linewidth]{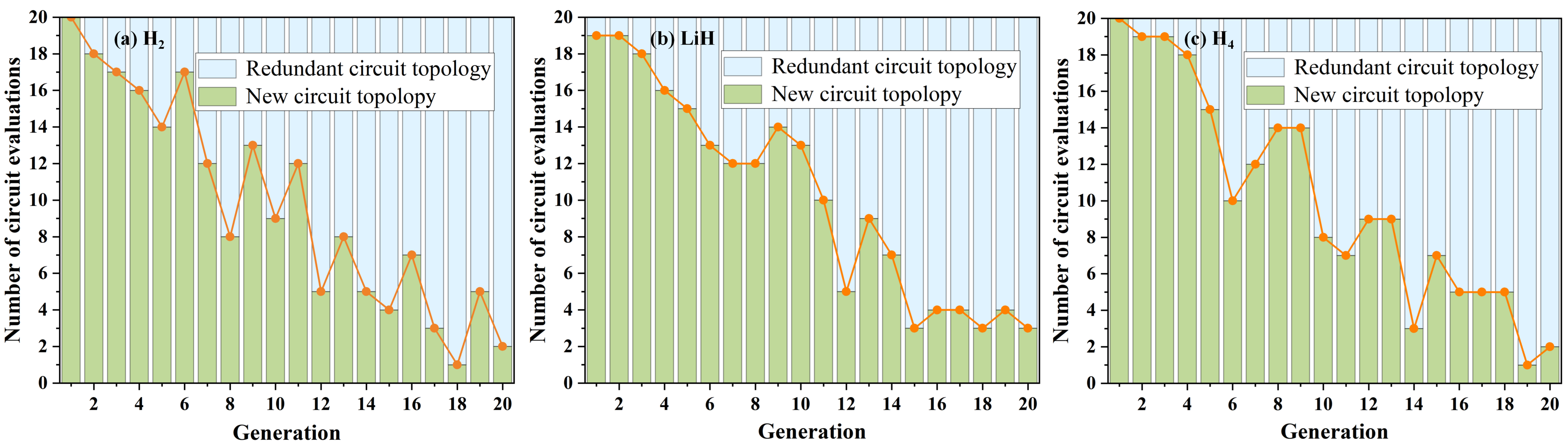}
    \caption{Generational evolution of the ZX topological reuse ratio. Panels (a), (b), and (c) show the results for $H_2$, $LiH$, and $H_4$, respectively. The population size is 20 in each generation. Green bars denote new circuit topologies that require evaluation, while blue bars denote redundant circuit topologies identified by the ZX topological reuse mechanism.}
    \label{FIG.5}
\end{figure*}

Figure \ref{FIG.5} shows the reuse of candidate topologies during the ZX-QAS search. The three molecular experiments exhibit a similar trend. New topologies dominate at the early stage, while the fraction of redundant topologies gradually increases as the evolution proceeds. This indicates that the evolutionary search does not keep sampling randomly over the whole search space. Instead, it gradually concentrates on several structurally favorable neighborhoods. Within these neighborhoods, candidate circuits may have different encodings, but they show similar topological features after ZX simplification. The topological reuse mechanism can therefore reduce repeated evaluations.

For $H_2$ (Fig. \ref{FIG.5}(a)), almost all candidate circuits require evaluation in the first few generations. This indicates that the initial population has high structural diversity. The number of real evaluations then decreases steadily and reaches its minimum value of 1 at generation 18. This means that most candidates are identified as low-novelty structures at the later stage. This behavior is consistent with the smaller search space of $H_2$, where candidate structures are more likely to recur.

For $LiH$ (Fig. \ref{FIG.5}(b)), the decrease in evaluations is smoother. New topologies still dominate during the first five generations. The number of evaluations then gradually decreases to about 3-5 per generation. Compared with $H_2$, $LiH$ shows smaller fluctuations in the middle stage. This suggests a more stable search process, in which the database of evaluated ZX features gradually covers the main structural neighborhoods. At the later stage, most candidates are handled by the reuse mechanism. This demonstrates that ZX topological reuse can still reduce evaluation cost in a more complex molecular task.

For $H_4$ (Fig. \ref{FIG.5}(c)), the number of evaluations is also close to 20 per generation at the early stage, but it decreases clearly from generation 6. A short increase appears around generations 8-13. This suggests that the evolutionary search enters new structural regions during this stage, so additional evaluations are needed to update the cache and the feature database. At the later stage, the number of evaluations decreases again to about 1-5 per generation. This indicates that the algorithm has shifted from exploring new structures to reusing similar ones. Such fluctuations are consistent with the larger hydrogen-chain system. Its search space is larger, and transitions between structural clusters occur more frequently. Therefore, the reuse ratio does not increase monotonically.

Figure \ref{FIG.5} shows that the ZX topological reuse mechanism does not skip evaluations aggressively at the beginning of the search. It first builds reliable caches through evaluations and then gradually increases the reuse ratio at later generations. This design preserves early-stage exploration and reduces repeated evaluations after the search has converged toward local structural clusters. The results demonstrate that the ZX-calculus-based topological reuse mechanism can convert structural similarity between candidate circuits into actual savings in evaluation cost. Based on the experimental results shown in Figures \ref{FIG.3} and \ref{FIG.4}, it shows that ZX-QAS significantly reduces the number of evaluations while maintaining an acceptable energy error. This achieves a practical cost-accuracy tradeoff for QAS.

\begin{table*}
%\begin{minipage}{100pt}
\caption{Cache hits and reuse statistics of ZX-QAS. The table reports the number of exact architecture cache hits, the number of ZX similarity reuse hits, and their corresponding ratios for the $H_2$, $H_4$, and $LiH$ tasks under 400 EA samples. An \textit{exact architecture cache hit} means that the same candidate circuit is sampled again and its cached result is reused directly. A \textit{ZX similarity reuse hit} means that the candidate circuit is not identical to any previously evaluated circuit, but its ZX feature vector is sufficiently close to that of an evaluated architecture. The fitness of its nearest evaluated neighbor is therefore reused.}
\label{Table-1}
\begin{ruledtabular}
\begin{tabular}{ccccccc}
Task& Qubits& Samples& Exact architecture cache hits& ZX similarity reuse hits& Exact cache hit rate&ZX similarity reuse rate\\ 
\hline
$H_2$& 4& 400& 44& 160& $11\%$&$40\%$\\
$H_4$& 8& 400& 5& 193& $1.25\%$&$48.25\%$\\
$LiH$& 6& 400& 7& 190& $1.75\%$&$47.50\%$\\
\end{tabular}
\end{ruledtabular}
%\end{minipage}
\end{table*}

Table \ref{Table-1} shows the main source of evaluation-cost reduction in ZX-QAS for molecular ground-state energy estimation. For the $H_2$ task, there are 44 exact architecture cache hits, corresponding to a hit rate of $11\%$. There are also 160 ZX similarity reuse hits, corresponding to a reuse rate of $40\%$. This indicates that, in a smaller search space, identical architectures are sampled with a relatively high probability. Exact caching can therefore already save part of the evaluation cost. In addition, ZX similarity reuse identifies many low-novelty candidates and increases the total reuse ratio to $51\%$.

For the $H_4$ and $LiH$ tasks, the exact architecture cache hit rates decrease sharply to only $1.25\%$ and $1.75\%$, respectively. As the number of qubits increases, the search space expands rapidly, and the probability of sampling exactly the same circuit during evolution becomes much lower. Therefore, exact architecture caching alone provides little effective evaluation saving. In contrast, ZX similarity reuse reaches $48.25\%$ for $H_4$ and $47.50\%$ for $LiH$. It becomes the main source of evaluation-cost reduction.

Table \ref{Table-1} shows that the advantage of ZX-QAS comes from using structural features after ZX simplification to identify low-novelty candidate circuits. In larger search spaces, such as those of $H_4$ and LiH, repeated sampling rarely occurs, but structurally similar candidate circuits are still common. The ZX similarity reuse rate remains at about $47\%$-$48\%$, which results from the identification and reuse of similar topological structures.

\subsection{One-dimensional transverse-field Ising model}
The one-dimensional transverse-field Ising model (1D TFIM) is a standard model for studying quantum many-body systems, quantum phase transitions, and entanglement structures. Its Hamiltonian is usually written as
\begin{equation}
H_{\rm TFIM}=-J\sum_{i=1}^{N-1} Z_i Z_{i+1}-h\sum_{i=1}^{N} X_i .
\label{Eq.20}
\end{equation}

where \(J\) is the interaction strength between neighboring spins, and \(h\) is the transverse-field strength. When \(J\) and \(h\) are comparable, the system lies in a regime with strong quantum fluctuations, and its ground state contains nontrivial many-body correlations. In our implementation, open boundary conditions are used. Therefore, the nearest-neighbor interaction term runs from \(i=1\) to \(N-1\), and no periodic boundary term \(Z_NZ_1\) is included. Therefore, TFIM can test not only the expressibility of variational circuits for quantum many-body ground states, but also the generalization ability of a quantum architecture search method on non-molecular Hamiltonians.

In this work, we use an \(N=8\) qubit 1D TFIM as the many-body benchmark, with \(J=1.0\) and \(h=1.0\). This parameter setting is close to the quantum critical region and provides a challenging search problem. The exact ground-state energy is obtained by sparse diagonalization of the Hamiltonian matrix and is used as the reference value for error calculation. The number of supernet experts is set to \(K=5\), and the circuit depth is set as \(l_{\min}=3\) and \(l_{\max}=10\). For the ZX topological reuse mechanism, the similarity threshold is set to \(\tau=0.23\), the maximum redundancy-penalty coefficient to \(\gamma_{\max}=0.035\), and the minimum real-evaluation probability to \(p_{\min}=0.12\).

\begin{figure}[htbp]
    \centering
    {\includegraphics[width=0.35\textwidth]{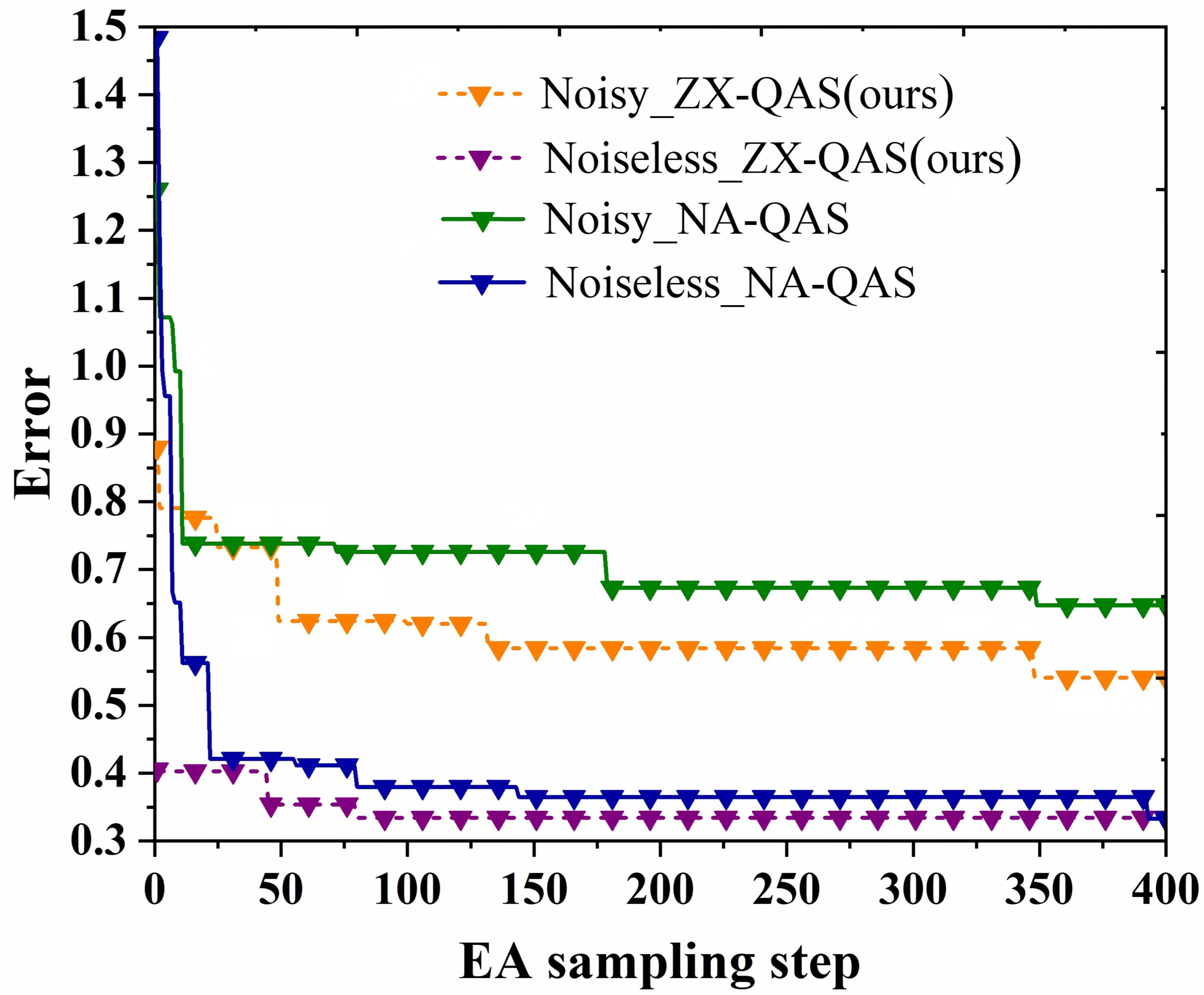}}

    \caption{Error convergence curves for the 1D TFIM. The horizontal axis denotes the evaluation step, and the vertical axis denotes the error relative to the exact ground-state energy. The orange and purple curves show the errors of the proposed ZX-QAS under noisy and noiseless conditions, respectively. The green and blue curves show the errors of NA-QAS without ternary Gray-code implicit mapping and ZX topological reuse under noisy and noiseless conditions, respectively.}
    \label{FIG.6}
\end{figure}

Figure 6 shows the error convergence behavior for the TFIM task. Under noisy conditions, the error of ZX-QAS decreases rapidly from about 0.86 and stabilizes near 0.54 at the later stage of the search. Here the error is measured relative to the exact ground-state energy of the dimensionless TFIM Hamiltonian with \(J=h=1\).In comparison, NA-QAS finally remains near 0.65. This shows that, for this many-body model, introducing ZX topological reuse does not degrade the search quality. Instead, it leads to a lower final error under a limited evaluation budget.

Under noiseless conditions, both methods achieve substantially lower errors than in the noisy setting. ZX-QAS reaches a final error of about 0.33, which is slightly lower than that of NA-QAS. This result indicates that noise remains the main factor limiting the accuracy of TFIM ground-state search. The structural reuse mechanism in ZX-QAS maintains a stable search direction under both noiseless and noisy conditions.

\begin{figure}[htbp]
    \centering
    {\includegraphics[width=0.35\textwidth]{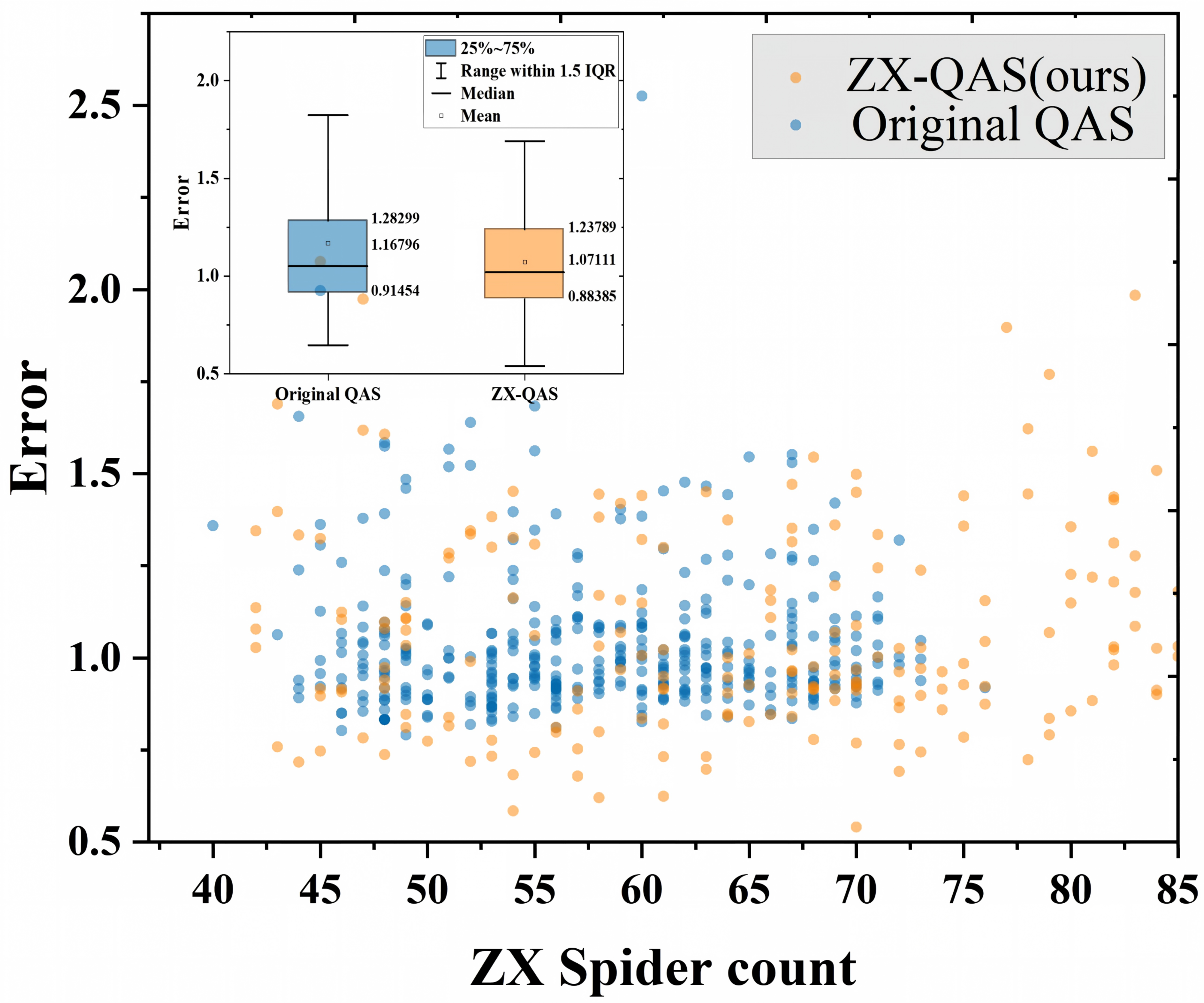}}

    \caption{Relation between ZX topological complexity and energy error in the 1D TFIM. The main plot shows the scatter distribution of candidate-circuit errors as a function of the spider count after ZX simplification. The inset shows box plots of the error distributions for ZX-QAS and Original QAS. Orange points denote ZX-QAS, while blue points denote Original QAS without ternary Gray-code implicit mapping and ZX topological reuse.}
    \label{FIG.7}
\end{figure}

Figure \ref{FIG.7} further shows the error distribution of candidate circuits. The scatter plot shows that the error is not a simple monotonic function of the ZX spider count. Instead, it has a broad distribution within a certain range of topological complexity. This indicates that the spider count alone cannot determine circuit performance. It is therefore more suitable as a regularization term for structural complexity than as the only screening criterion. The box plots show that the interquartile range of ZX-QAS is slightly lower than that of Original QAS, and that high-error outliers are also reduced. Although the two methods have similar median errors, ZX-QAS preserves more candidates in the low-error region and compresses the upper tail of the error distribution. This suggests that ZX topological reuse does not simply discard candidates. Instead, it uses structural similarity to guide the search toward more stable topological regions.

\begin{figure}[htbp]
    \centering
    {\includegraphics[width=0.35\textwidth]{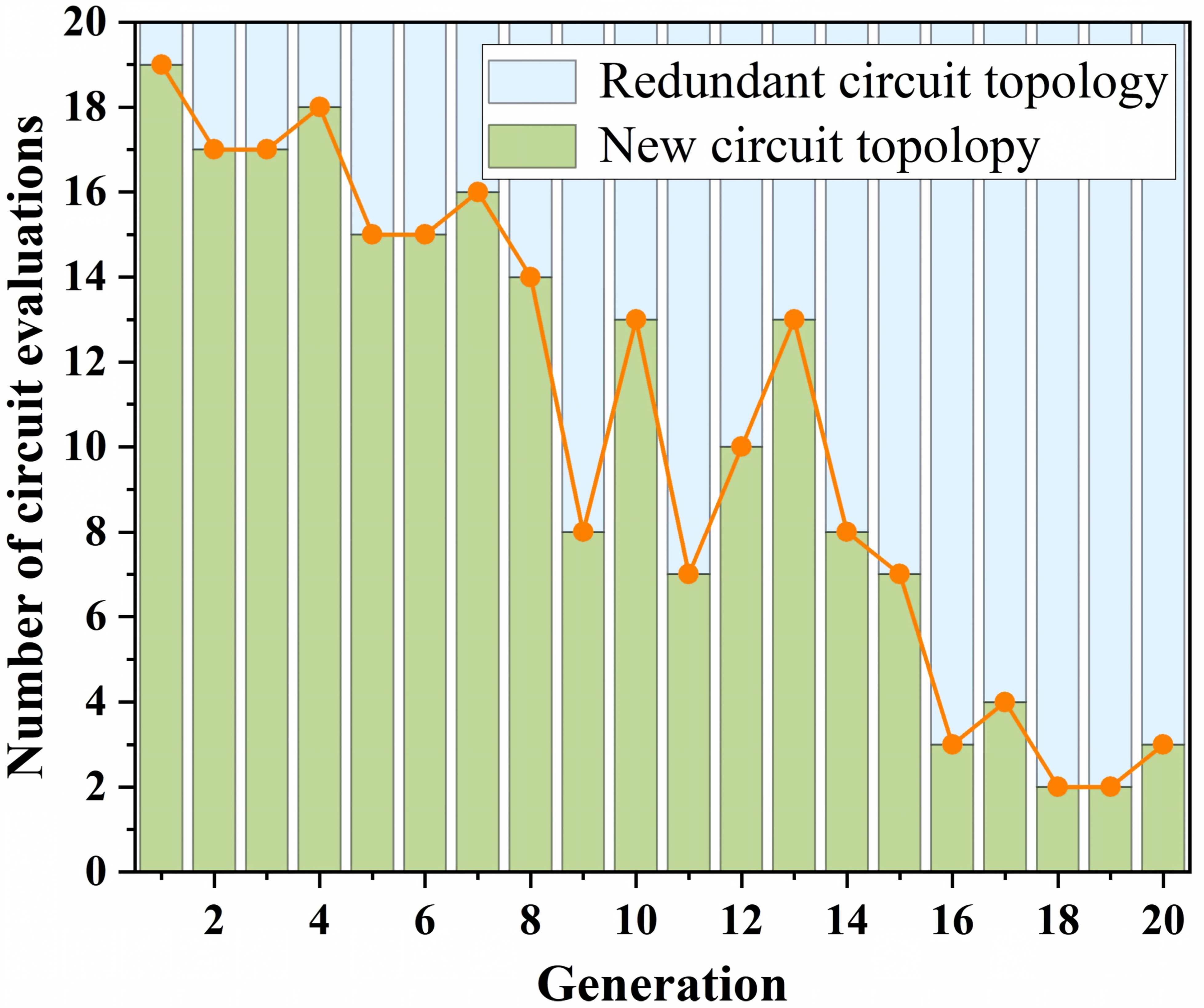}}

    \caption{Generational evolution of the ZX topological reuse ratio during the search for the 1D TFIM. The population size is 20 in each generation. Green bars denote new circuit topologies that require evaluation, while light-blue bars denote redundant circuit topologies identified by the ZX topological reuse mechanism.}
    \label{FIG.8}
\end{figure}

Figure \ref{FIG.8} shows the numbers of evaluated candidates and redundant-topology candidates in each generation. In the first few generations, the number of evaluations is close to 20 per generation. This indicates that the initial population mainly explores new topologies. As the evolution proceeds, the number of evaluations gradually decreases and often drops to about 2-4 per generation at the later stage. This shows that the population gradually concentrates on several neighborhoods with similar structures. The ZX topological reuse mechanism can identify these low-novelty candidates and reduce repeated evaluations. A short increase appears around generations 10-13. This indicates that the algorithm is not completely locked by the early cache. When it enters a new structural region, it still preserves evaluations. This behavior is consistent with the design of the minimum real-evaluation probability, which helps reduce the risk of incorrect reuse.

In the 1D TFIM experiment, the evolutionary algorithm performs 400 samples in total. Among them, the exact architecture cache is hit 6 times, corresponding to a hit rate of $1.5\%$. ZX similarity reuse is triggered 183 times, corresponding to a reuse rate of $45.75\%$.

ZX-QAS achieves a cost-accuracy tradeoff in the TFIM task. The error convergence curves in Fig. \ref{FIG.6} show that ZX-QAS reaches a lower final error than the baseline under noisy conditions. Fig. \ref{FIG.7} shows that ZX topological features are statistically correlated with candidate performance and can serve as structural priors in the evaluation stage. Fig. \ref{FIG.8} shows that the number of evaluations decreases substantially during the evolutionary process. These results indicate that the ZX topological reuse mechanism is not limited to molecular ground-state energy estimation. It can also be extended to the 1D TFIM, which contains many-body correlations. The ZX topological reuse mechanism therefore turns circuit simplification into a tool for evaluation-budget management in QAS. It reduces the number of expensive evaluations while maintaining an acceptable energy error.

\subsection{Ablation analysis of the Gray-code implicit mapping}
The previous section shows that ZX-QAS benefits from structure-aware evaluation. Here, we isolate the role of the ternary Gray-code implicit mapping. We test whether this encoding reduces storage overhead and preserves local circuit changes in the discrete \textit{ansatz} space. 

To verify the effect of this mechanism, we design two independent ablation experiments. The first experiment compares the number of physical gate changes induced by adjacent index changes under ternary Gray-code mapping and standard ternary mapping. We use \(N=8\) qubits, enumerate all \(3^8\) single-qubit rotation-gate combinations, and calculate the Hamming distance \(D_H\) between the gate sequences corresponding to two adjacent indices \(u\) and \(u+1\). Here, \(D_H\) denotes the number of single-qubit rotation gates that change simultaneously in one adjacent encoding step.

The second experiment compares the memory footprint of explicit Cartesian storage with that of the proposed on-demand decoding strategy. According to the search-space definition in Sec. III, each layer contains \(3^N\) rotation-gate combinations and \(|\rm{CNOT}_{\rm space}|=1+\frac{N(N-1)}{2}\) sparse entangling choices. All memory results are reported in GB.

\begin{figure}
    \centering
    {\includegraphics[width=0.35\textwidth]{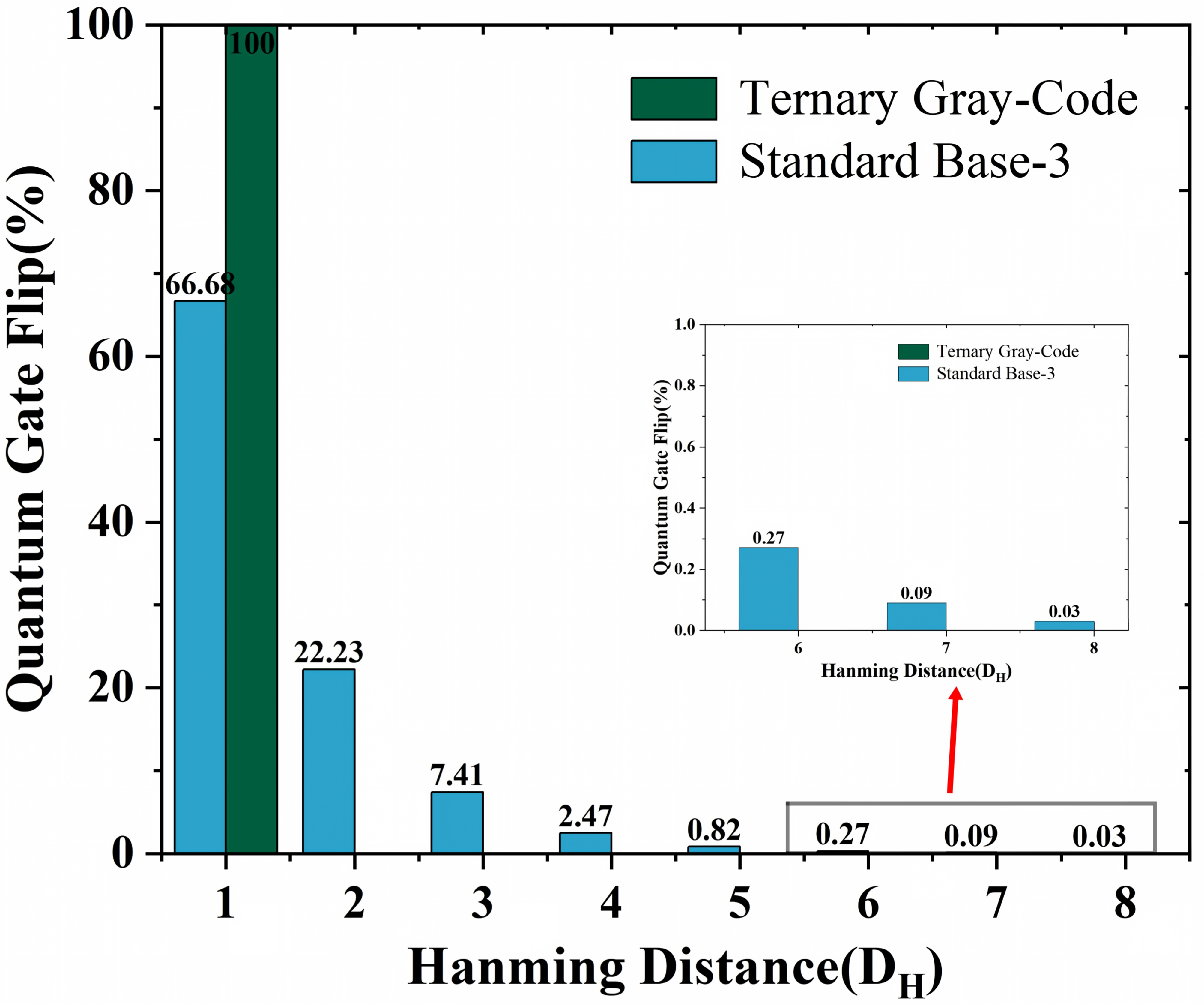}}

    \caption{Distribution of local gate changes for ternary Gray-code mapping and standard ternary mapping. The horizontal axis denotes the Hamming distance \(D_H\) between the gate sequences corresponding to adjacent encodings, and the vertical axis denotes the occurrence probability of each distance. Green bars denote ternary Gray-code mapping, while blue bars denote standard ternary mapping. }
    \label{FIG.9}
\end{figure}

Figure \ref{FIG.9} shows that ternary Gray-code mapping gives \(D_H=1\) for all adjacent index changes. In other words, each adjacent encoding change corresponds to the change of only one single-qubit rotation gate. In contrast, under standard ternary encoding, only about $66.68\%$ of adjacent changes correspond to a single-gate change. About $22.23\%$ change two gates simultaneously, about $7.41\%$ change three gates simultaneously, and some cases still change six, seven, or even eight gates at the same time. Although these high-Hamming-distance events have small probabilities, they correspond to abrupt structural changes of candidate circuits during evolutionary search. A local perturbation in the encoding can therefore become a simultaneous jump of multiple physical gates.

This result shows that Gray-code mapping does not change the expressibility of the search space itself. Instead, it gives candidate circuits stronger local continuity. As a result, genetic mutation or a local search step is more likely to correspond to a local change in the circuit structure. Gray-code mapping therefore helps the search process preserve locally useful structures from existing candidates. This observation is consistent with the results in Secs. IV B and IV C. In both molecular ground-state energy estimations and the 1D TFIM experiment, ZX-QAS maintains stable search behavior under a limited evaluation budget.

\begin{figure}
    \centering
    {\includegraphics[width=0.35\textwidth]{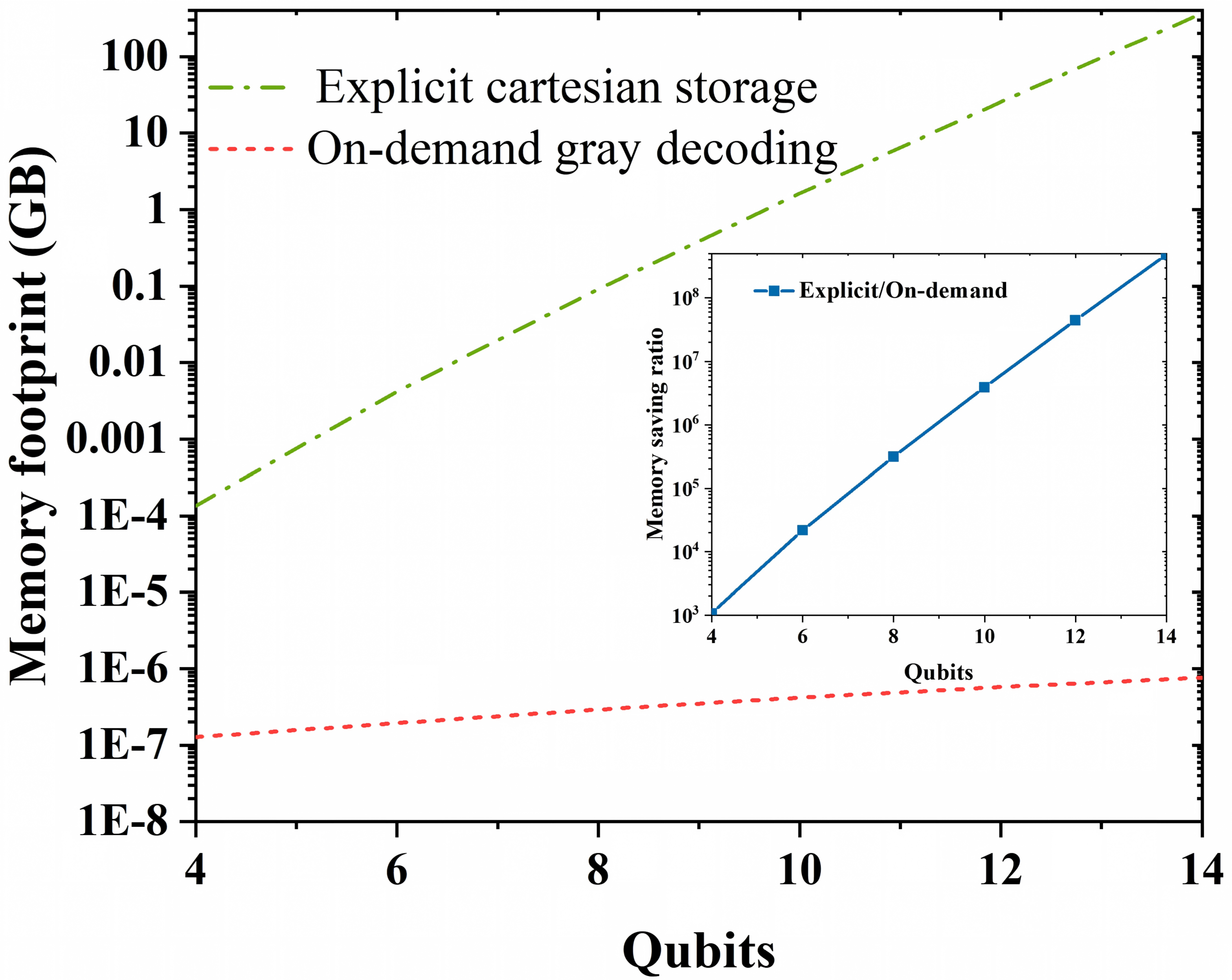}}
    
    \caption{Comparison of memory footprints between explicit Cartesian storage and on-demand Gray-code decoding. The horizontal axis denotes the number of qubits, and the vertical axis denotes the memory footprint in GB on a logarithmic scale. The green curve shows the memory required to explicitly store the Cartesian product of the rotation-gate and CNOT subspaces. The red curve shows the memory required by on-demand Gray-code decoding. The inset shows the memory saving ratio.}
    \label{FIG.10}
\end{figure}

Figure \ref{FIG.10} further shows the memory advantage of the implicit mapping. As the number of qubits increases, the memory requirement of explicit Cartesian storage grows rapidly. For example, when \(N=14\), explicitly storing the combinations of the single-layer rotation-gate subspace and the sparse CNOT subspace already requires about \textbf{367} GB of memory. In contrast, on-demand Gray-code decoding only needs to store the current candidate encoding and the CNOT lookup table. Its memory footprint therefore remains very small. The inset shows that the memory saving ratio over explicit storage increases rapidly with the number of qubits and reaches above the order of \(10^8\) for 14 qubits.

This result shows that the scalability of the proposed method does not rely on generating a complete library of candidate circuits in advance. The explicit Cartesian method must construct and store all possible layer structures before the search begins. In contrast, the proposed method dynamically decodes the rotation gates and entangling gates from an integer index only when a candidate is evaluated. Thus, the search space still retains exponential expressibility in its mathematical definition, while the actual storage cost is decoupled from the size of the full candidate space. This is especially important for VQE tasks such as molecular systems and TFIM. As the number of qubits and the maximum circuit depth increase, the practical bottleneck of QAS is often not whether a single circuit can be trained, but whether candidate structures can be continuously generated, evaluated, and compared under limited memory and evaluation budgets.

In summary, the ternary Gray code is not merely an encoding trick. It is a structural mapping that connects discrete search steps with local quantum-gate changes. On-demand decoding further avoids the storage explosion caused by explicit construction of the search space.

\section{Conclusion}
We propose a noise-aware quantum architecture search framework based on ZX-calculus topological reuse (ZX-QAS) to reduce repeated evaluations of candidate \textit{ansatz}es under NISQ conditions. First, we construct an implicit search space using a ternary Gray code. This encoding makes adjacent codes correspond to local single-qubit gate changes and avoids out-of-memory problems caused by explicitly storing the Cartesian product of the rotation-gate and entangling-gate subspaces. Second, we apply ZX topological reuse in the fitness-evaluation stage. The method identifies low-novelty candidates through distances between ZX similarity features and reduces the evaluation cost by probabilistic reuse. The experiments show that ZX-QAS maintains stable convergence in molecular ground-state energy estimations and the 1D TFIM. It also reduces the number of evaluations substantially, with evaluation-cost savings above 40\%.

It should be noted that ZX-QAS is designed as an accuracy-cost tradeoff under hardware and noise constraints. The topological reuse mechanism may introduce approximation errors. Therefore, the similarity threshold, reuse penalty, and minimum real-evaluation probability should be adjusted according to the task scale. Future work may introduce a more refined learning-based ZX structural distance. Overall, our results show that ZX-QAS can use ZX graphs as structural priors in QAS. This provides a feasible route toward low-cost and noise-aware \textit{ansatz} design on NISQ devices.

\begin{acknowledgments}
This work was supported by the National Natural Science Foundation of China (No. 62371238), and supported by the Key Project of Science Foundation of Jiangsu Province (No. BK20243046).
\end{acknowledgments}

\bigskip
\textbf{DATA AVAILABILITY}
\par
The data that support the findings of this article are openly available \cite{web_misc}, embargo periods may apply

\bigskip
\textbf{ORCID iDs}\\
Hui Zeng https://orcid.org/0000-0002-7657-6714\\
Dazhi Ding https://orcid.org/0000-0001-8522-6233\\

\newpage
\bibliography{reference}% Produces the bibliography via BibTeX.

\end{document}